\documentclass[aps, prl, twocolumn, amsfonts, floatfix]{revtex4}
\usepackage{graphicx}
\usepackage{amsmath,amsfonts}
\usepackage{epsfig,color}

\begin{document}
\newcommand{\be}{\begin{eqnarray}}
\newcommand{\ee}{\end{eqnarray}}
\newcommand\del{\partial}
\newcommand\nn{\nonumber}
\newcommand{\Tr}{{\rm Tr}}
\newcommand{\Str}{{\rm Trg}}
\newcommand{\mat}{\left ( \begin{array}{cc}}
\newcommand{\emat}{\end{array} \right )}
\newcommand{\vect}{\left ( \begin{array}{c}}
\newcommand{\evect}{\end{array} \right )}
\newcommand{\tr}{{\rm Tr}}
\newcommand{\hm}{\hat m}
\newcommand{\ha}{\hat a}
\newcommand{\hz}{\hat z}
\newcommand{\hze}{\hat \zeta}
\newcommand{\hx}{\hat x}
\newcommand{\hy}{\hat y}
\newcommand{\tm}{\tilde{m}}
\newcommand{\ta}{\tilde{a}}
\newcommand{\U}{\rm U}
\newcommand{\diag}{{\rm diag}}
\newcommand{\tz}{\tilde{z}}
\newcommand{\tx}{\tilde{x}}
\newcommand{\arcsinh}{{\rm arcsinh}}
\definecolor{red}{rgb}{1.00, 0.00, 0.00}
\newcommand{\rd}{\color{red}}
\definecolor{blue}{rgb}{0.00, 0.00, 1.00}
\definecolor{green}{rgb}{0.10, 1.00, .10}
\newcommand{\blu}{\color{blue}}
\newcommand{\green}{\color{green}}

\title{The Dirac spectrum in Complex Langevin Simulations of QCD}

\author{K. Splittorff}
\affiliation{Discovery Center, The Niels Bohr Institute, University of Copenhagen, 
Blegdamsvej 17, DK-2100, Copenhagen {\O}, Denmark}

\date   {\today}
\begin  {abstract}
We show that the spectrum of the Dirac operator in complex Langevin simulations of QCD 
at non-zero chemical potential must behave in a way which is radically different
from the one in simulations with ordinary non-complexified gauge fields:
At low temperatures the small eigenvalues of the Dirac operator must be 
inside the quark mass for chemical potentials as large as a third of the nucleon mass. 
In particular, in the chiral limit the Dirac eigenvalues of complex Langevin simulations must 
accumulate at the origin. 
\end{abstract}
\maketitle

\section{Introduction}

Complex Langevin dynamics \cite{Klauder,Parisi} is the most promising and 
challenging approach to the sign problem in QCD at non-zero chemical potential.
It is promising because it can solve sign problems that are exponentially hard 
in the volume \cite{AartsBose,AS}. It is challenging because such 
a success is not guaranteed \cite{AFP,AartsXY,MS}. See \cite{AartsRev} for a review. 

Before we get into the details, let us recall why a first principles non-perturbative 
computation is necessary in order to determine the QCD phase diagram. The phase 
structures of strongly interacting matter are intimately 
linked to the transition between the hardronic phase and phases where quark and gluons are    
the natural degrees of freedom. The values of the chemical potential and temperatures
at the transition are therefore on the hadronic scale for which QCD is highly non-perturbative.
Hence the natural tool to apply is lattice QCD  \cite{PhilippeRev}. For zero chemical potential the fermion determinant 
is real and lattice QCD with two mass degenerate flavors is effectively studied using Monte Carlo sampling.
One important message to take from these simulations is that the transition from hadronic matter 
is highly sensitive to the quark mass. A first principle non-perturbative dynamical computation is therefore essential.  

At non-zero chemical potential it is natural to use complex Langevin dynamics in lattice QCD 
since the fermion determinant is complex. In this approach the gauge fields are complexified and 
one uses the complex valued action to define the drift of the gauge field configurations with Langevin 
time. Observables are evaluated on the trajectory traced out by the gauge fields and the expectation 
value is simply the average for sufficiently many Langevin time steps.

During the past year improved numerical methods \cite{GaugeCooling} have allowed to simulate full QCD at non-zero 
chemical potential, $\mu$, with complex Langevin in different parameter regions \cite{Sexty:2013ica,all-orders}. 
The key challenge at present is to understand if complex Langevin also allows us to simulate QCD 
at low temperature, small quark mass and chemical potential up to a third of the nucleon mass. 
This region is {\sl the test ground} for numerical efforts to solve the QCD sign problem. From Monte Carlo simulations 
at $\mu=0$ we know that chiral symmetry is spontaneously broken, and hence the vacuum is 
dominated by pions. As pions are neutral wrt.~the quark charge no phase transitions are expected 
in the region $\mu\sim m_\pi/2$. However, once the 
chemical potential passes half the pion mass the sign problem becomes extremely severe, see 
e.g.~\cite{SVphase}, and unless the sign problem is under full control one is likely to see an
unphysical phase transition at $\mu=m_\pi/2$, see eg.~\cite{SplitRev}.  

The sign problem encountered beyond $\mu=m_\pi/2$ becomes particularly clear when considered 
from the perspective of the Dirac eigenvalues. In standard QCD without any complexification of the 
gauge fields the eigenvalues of the Dirac operator form a band along the imaginary axis \cite{TV,AOSV,SplitRev}. 
When the chemical potential increases to $m_\pi/2$ this band reaches the quark mass. 
For larger values of the chemical potential the eigenvalue density moves beyond the quark mass.
In a theory without a sign problem, such as phase quenched QCD, this will imply that the 
theory is in a new phase characterised by pion condensation \cite{SS,SSS,OSV-phaseSpec}. For full QCD, however, in a region 
extending from the quark mass and outward the eigenvalue density becomes complex valued 
and highly oscillatory \cite{AOSV}. See Fig.~\ref{fig:fixed-m} and \ref{fig:fixed-mu} upper panels. 
These oscillations wipe out the unphysical phase transition 
at $\mu=m_\pi/2$ \cite{OSV1,OSV2} and cause the discontinuity of the chiral condensate at zero 
quark mass.

In this paper we consider, for the first time, the general properties of the eigenvalues of the Dirac operator 
evaluated on the complexified gauge field configurations generated by the complex Langevin dynamics.
We show that the behaviour of the Dirac spectrum in complex Langevin simulations 
must be drastically different from that with real gauge fields: The real part of the small eigenvalues of the Dirac operator in complex Langevin 
must have a magnitude less than the quark mass. This holds for low temperature and chemical potentials as large as a third of the nucleon mass. 
In particular, in the chiral limit the small Dirac eigenvalues of complex Langevin simulations must 
accumulate at the origin. See the lower panels of Fig.~\ref{fig:fixed-m} and \ref{fig:fixed-mu}.
The accumulation of Dirac eigenvalues at the origin is in sharp contrast to the fact that the 
chemical potential breaks the anti-Hermiticity of the Dirac operator but, as we will show, it is not 
excluded by the standard arguments.

To set the stage we first recall the two different links between the spectrum of the Dirac operator with 
real gauge fields and the chiral condensate at zero \cite{BC} and non-zero chemical potential 
\cite{OSV1,OSV2}. We then turn to the spectrum of the complex Langevin Dirac operator at 
non-zero $\mu$. The 7 main points are highlighted by enumeration.

\section{Banks-Casher at $\mu=0$}

The successful application of importance sampling in lattice QCD for 2 degenerate quark flavors 
at $\mu=0$ relies on the fact that we (in this case) deal with a statistical system: The fermion determinant
is real and its square positive. 
One way to see this is to express the determinant in terms of the eigenvalues $i\lambda_k$ of the Dirac 
operator at $\mu=0$
\be
\det(D_{\mu=0}+m) & = & \prod_k(i\lambda_k+m)(-i\lambda_k+m)\nn\\
& = & \prod_k(\lambda_k^2+m^2),
\ee 
where we have used that the Dirac operator is anti-Hermitian and anti-commutes with $\gamma_5$.
For simplicity we do not consider topological zero modes. 

A central observable which gives us insights on the phases of strongly interacting matter is 
the chiral condensate, 
\be
\Sigma(m) & = & \left\langle \Tr \frac{1}{D_{\mu=0}+m}\right\rangle_{N_f,\mu=0}.
\ee
Here $\left\langle \ldots \right\rangle_{N_f,\mu=0}$ denotes the QCD expectation value with 
$N_f$ dynamical quark flavors of equal mass $m$ and zero chemical potential $\mu$. 
Expressed in terms of the Dirac eigenvalues we have
\be
\Sigma(m)  & = &   \left\langle \sum_k \frac{1}{i\lambda_k+m} \right\rangle_{N_f,\mu=0}. 
 \ee
Equivalently, we can write this as (for a detailed discussion see \cite{LS})
\be
\Sigma(m) & = & \int d\lambda \ \rho_{N_f}(\lambda,m)\frac{1}{i\lambda+m} 
\label{BCm}
\ee
where the eigenvalue density is defined by
\be
\rho_{N_f}(\lambda,m) = \left\langle \sum_k \delta(\lambda-\lambda_k) \right\rangle_{N_f,\mu=0}.
\ee
Since the non-zero eigenvalues come in pairs of opposite sign on the imaginary axis the integrand 
of (\ref{BCm}) forms a $\delta$-function in the chiral limit, $m\to 0$,
which singles out $\lambda=0$. This gives the Banks-Casher relation \cite{BC}
\be
\lim_{m\to0}\Sigma(m) = \frac{\pi}{V}\rho(0) .
\label{BCrel}
\ee
In words: at zero chemical potential the density of Dirac eigenvalues at the origin of the imaginary 
axis is proportional to the chiral condensate in the chiral limit.

\section{Non-complexified gauge fields at $\mu\neq0$}

Next, let us consider QCD at non-zero chemical potential, without 
any complexification of the gauge fields. The first thing to notice is that the 
standard Banks-Casher relation (\ref{BCrel}) is invalidated. Since the Dirac operator is 
no longer anti-Hermitian the eigenvalues, $z_k$, of the Dirac operator, $D_\mu$,
move into the complex plane. 
The chiral condensate,
\be
\Sigma(m;\mu) & = & \left\langle \Tr \frac{1}{D_\mu+m}\right\rangle_{N_f,\mu},
\ee
can still be expressed in terms of the eigenvalues $z_k$ of $D_\mu$
\be
\Sigma(m;\mu) & = &  \left\langle \sum_k \frac{1}{z_k+m} \right\rangle_{N_f,\mu} \\
 & = & \int d^2z \ \rho_{N_f}(z,m;\mu)\frac{1}{z+m} \nn
 \label{Sigma_mu_int}
\ee
where the eigenvalue density is
\be
\rho_{N_f}(z,m;\mu) = \left\langle \sum_k \delta^{(2)}(z-z_k) \right\rangle_{N_f,\mu}.
\label{def-rho-MC}
\ee
However, eventhough the non-zero eigenvalues still come in pairs of opposite sign, the integrand of 
(\ref{Sigma_mu_int}) does not form a $\delta$-function at the origin of the complex plane for $m\to0$. 	 
The standard Banks-Casher relation (\ref{BCrel}) is therefore not valid at non-zero $\mu$.

\begin{center}
\begin{figure}[t!]
\includegraphics[width=8cm,angle=0]{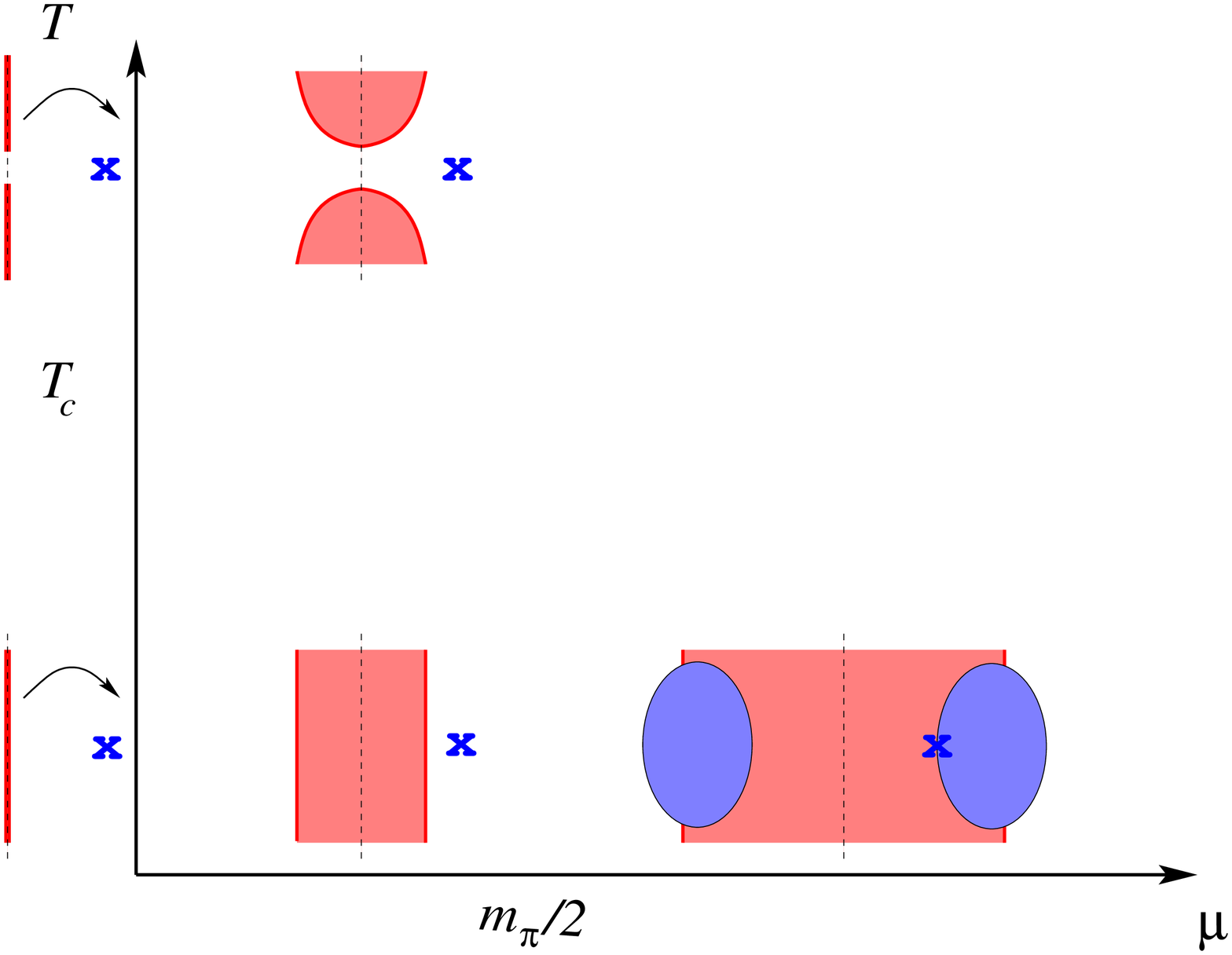}
\vspace{10mm}
\vfill
\includegraphics[width=8cm,angle=0]{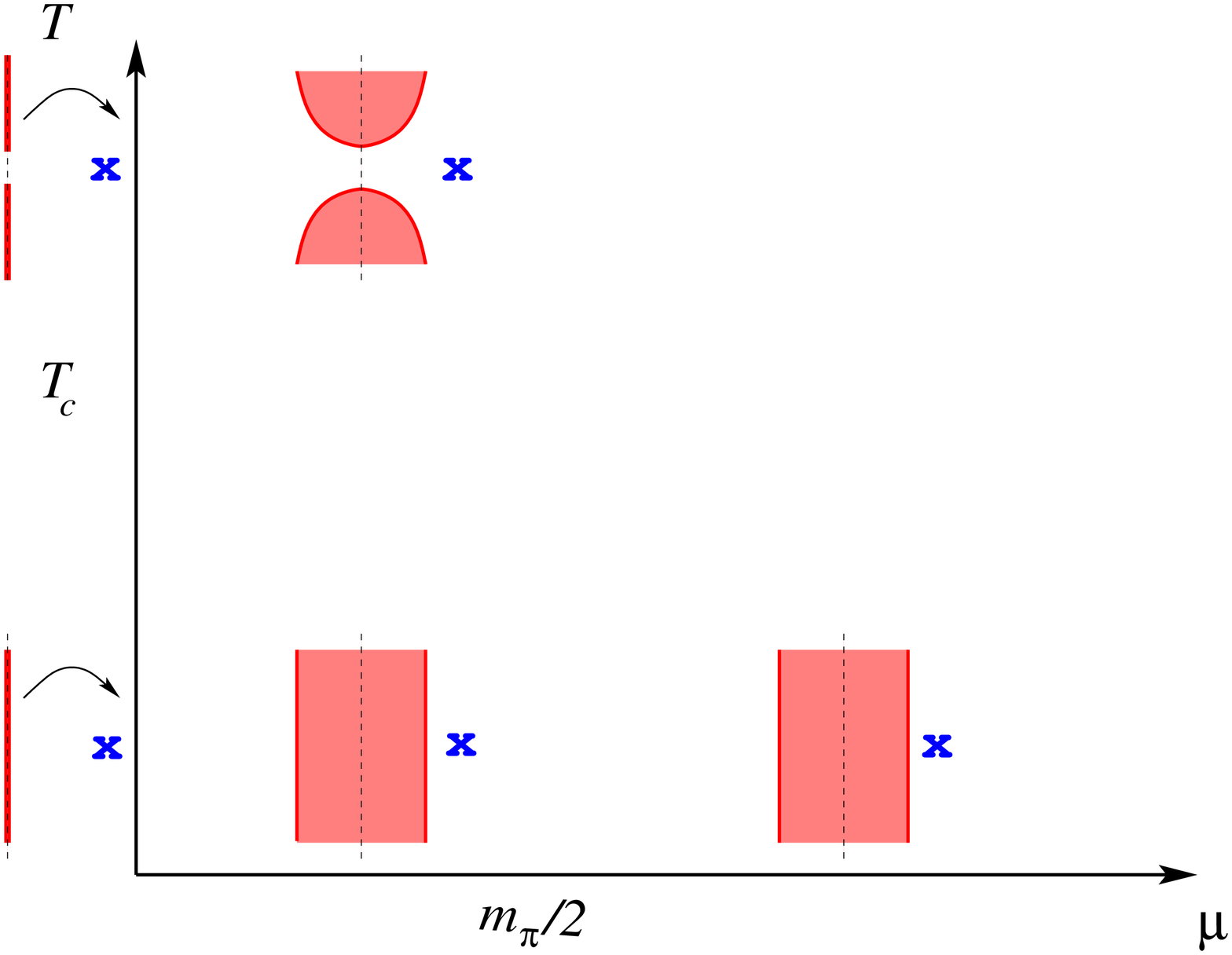}
\caption{\label{fig:fixed-m} Sketch of the low lying Dirac spectrum various places in the $(\mu,T)$-plane. The quark mass, illustrated by the blue cross, is fixed. {\bf Top:} For ordinary real gauge fields, {\bf Below:} in complex Langevin simulations. 
Within the red region the eigenvalue density is real and positive. The region where the eigenvalue density is complex valued and rapidly oscillating is marked in blue. This region is not possible to realise for complex Langevin, hence the eigenvalues must necessarily stay inside the quark mass even if $\mu>m_\pi/2$.}\vspace{5mm}
\end{figure}
\end{center}

What replaces the Banks-Casher relation is the fact that 
\begin{itemize}
\item[{\bf 1)}] $\rho_{N_f}(z,m;\mu)$ will take complex values for $\mu>m_\pi/2$ 
 \cite{AOSV}.
\item[{\bf 2)}] These complex 
valued oscillations of $\rho_{N_f}(z,m;\mu)$ are responsible for the formation of the chiral 
condensate \cite{OSV1,OSV2}.  
\end{itemize}
For illustrations, see the upper panels of Figure \ref{fig:fixed-m} and Figure \ref{fig:fixed-mu} as well as \cite{OSV-XQCD}.

It was possible to obtain these insights since the low lying eigenvalue
density can be determined uniquely from the flavor symmetries through chiral 
random matrix theory \cite{O} and chiral perturbation theory \cite{AOSV}.
The fact that the eigenvalue density takes complex values for $\mu>m_\pi/2$
is possible because the $\delta$-function in (\ref{def-rho-MC}) is weighted by 
the complex valued fermion determinant. The contribution from this oscillating 
part of the eigenvalue density to the integral in (\ref{Sigma_mu_int}) increases 
as the quark mass approaches zero and it changes sign as the quark mass passes through 
the origin. Hence, the complex and oscillating part of the 
spectral density is responsible for the spontaneous breaking of chiral symmetry \cite{OSV1,OSV2}.
Moreover, the oscillations ensures that $\Sigma(m;\mu)$ is smooth at $\mu=m_\pi/2$.

\subsubsection{Phase quenched QCD}

In contrast, for phase quenched theory QCD, where the fermion determinant is 
replaced by its absolute value, there is no sign problem and the eigenvalue density 
is necessarily real and positive. It forms a band along the imaginary axis of width 
$2\mu^2F_\pi^2/\Sigma$, independent of the quark mass \cite{TV,OSV-phaseSpec}. From this we learn that 
\begin{itemize}
\item[{\bf 3)}]
a positive density of eigenvalues in a band along the imaginary axis, 
leads to a vanishing value of the chiral condensate in the chiral limit, for any non-zero 
value of $\mu$. 
\item[{\bf 4)}]
a phase transition occurs when the quark mass hits this positive eigenvalue density.
\end{itemize}
The quark mass hits the strip of eigenvalues at $\mu=m_\pi/2$ and 
the new phase for $\mu>m_\pi/2$ is characterized by a non-zero pion condensate \cite{SS}\footnote{A real and 
positive eigenvalue density which forms a strip in the complex plane can lead to a discontinuity of the chiral condensate at $m=0$ 
if $\rho(z,m;\mu)\neq\rho(-z,m;\mu)$ \cite{KSV}.}.

\section{Complexified gauge fields at $\mu\neq0$}

Now let us turn to the complex Langevin approach.
Given an action $S({\bf x})$ which takes complex values when evaluated on the 
real fields ${\bf x}=(x_1,\ldots,x_N)$ the standard form of the complex Langevin 
drift equations, obtained by complexifying all fields $x_k\to x_k+iy_k$, is 
\cite{Parisi,Klauder,AartsRev}
\be
x^{(t+1)}_{k}&=&x^{(t)}_{k}-{\rm Re}\left[\frac{dS}{dx_k}\right]_{\bf{x}=\bf{x}^{(t)}+i\bf{y}^{(t)}}dt+\eta\sqrt{dt} \nonumber \\
y^{(t+1)}_{k}&=&y^{(t)}_{k}-{\rm Im}\left[\frac{dS}{dx_k}\right]_{\bf{x}=\bf{x}^{(t)}+i\bf{y}^{(t)}}dt .
\ee
Here $\eta$ is the Langevin noise term, see e.g.~\cite{AartsRev}.
The suberscript $t=1,\ldots,T$ refers to the Langevin time and expectation values of an 
observable $\cal{O}({\bf x})$ are measured simply by averaging over the observable 
evaluated on the Langevin trajectory for sufficiently large $T$
\be
\left\langle{\cal O}\right\rangle =\frac{1}{T}\sum_{t=1}^T {\cal O}(x_k^{(t)}+iy_k^{(t)}).
\ee
In applications to QCD at $\mu\neq0$ the fermions are integrated out and a trajectory of complexified gauge field configurations, 
$A_\nu^{(t)}$, is obtained from the complex Langevin drift equations \cite{Sexty:2013ica}.

The Dirac eigenvalues, $z^{(t)}_k$, on the Langevin trajectory of complexified gauge fields, $A_\nu^{(t)}$, are determined 
through the eigenvalue equation
\be
D_\mu(A_\nu^{(t)})\psi_k = z^{(t)}_k\psi_k .
\ee
The general properties of the eigenvalues, $z^{(t)}_k$, are much the same as with real gauge fields and non-zero 
chemical potential: The anti-commutation of the Dirac operator, $D_\mu(A_\nu^{(t)})$, with $\gamma_5$ is conserved 
but the anti-Hermiticity is broken. The non-zero eigenvalues, $z^{(t)}_k$, therefore come in pairs of opposite sign in the complex plane. 
\vspace{2mm}

A natural question to ask at this point is: {\sl can complex Langevin realize the formation of the chiral condensate through the 
oscillating eigenvalue density as in the ordinary approach to QCD}. The answer is: {\sl No}. 
\vspace{2mm}

To see why, let us now consider the formation of the chiral condensate in complex Langevin simulations of QCD. 
The Langevin process will generate a trajectory with complexified gauge field configurations, $A_\nu^{(t)}$, at Langevin 
time $t$ and the chiral condensate measured is  
\be
\Sigma^{(CL)}(m) & = & \frac{1}{T} \sum_t \Tr \frac{1}{D_\mu(A_\nu^{(t)})+m},
\ee
Expressed in terms of the Dirac eigenvalues we have 
\be
\Sigma^{(CL)}(m) = \frac{1}{T} \sum_t \sum_k \frac{1}{z^{(t)}_k+m}. 
\ee
Analogously to the standard case above let us rewrite this as
\be
\Sigma(m) = \int d^2z \; \frac{\rho^{(CL)}_{N_f}(z,m;\mu)}{z+m},
\ee
where 
\be
\rho^{(CL)}_{N_f}(z,m;\mu) = \frac{1}{T} \sum_t \sum_k \delta^{(2)}(z-z^{(t)}_k).
\ee
While this at first appears to be similar to the density for real gauge fields, (\ref{def-rho-MC}), there is one key 
difference  
\begin{itemize}
\item[{\bf 5)}] $\rho^{(CL)}_{N_f}(z,m;\mu)$ is by definition 
real and positive. 
\end{itemize}
Therefore, it is not possible to realise the formation of the chiral condensate through extreme and complex valued 
oscillations of the eigenvalue density.
\vspace{2mm}

 A next natural question is therefore: {\sl Does this imply that the complex 
Langevin approach fails for $\mu>m_\pi/2$?} The answer is: {\sl No, not nessesarely}.
\vspace{2mm}

 Eventhough $\rho^{(CL)}_{N_f}(z,m;\mu)$ is positive, the unphysical phase transition is avoided provided that 
\begin{itemize}
\item[{\bf 6)}] the density of the small eigenvalues must be located within a band bounded by 
$|{\rm Re}[z^{(t)}_k]|<|m|$ in the vicinity of the real axis. 
\end{itemize}
If not there will be a phase transition at the point where the quark mass hits the boundary of the eigenvalues.  In particular, 
\begin{itemize}
\item[{\bf 7)}] 
in the chiral limit the small eigenvalues of the Dirac operator 
evaluated on the complexified configurations must accumulate to a diverging density at the 
origin. 
\end{itemize}
In other words: Despite the fact that the chemical potential breaks the anti-Hermiticity of the Dirac operator, the  
small eigenvalues of the Dirac operator must be pushed to the origin by the decreasing value of the quark mass.
This is exemplified in the appendix.

The next question is to ask is, of course, {\sl is this possible?}

A natural way to get eigenvalues with small real part is from an anti-Hermitian operator.
However, since the gauge group of QCD is SU(3), it is not possible to choose a complexified gauge field configuration  
for which the Dirac operator at non-zero chemical potential is anti-Hermitian. To see this explicitly let us write the 
Hermitian and anti-Hermitian part of the complexified gauge fields, $A^{(t)}_\nu$ as $A_\nu^{(t)} = X_\nu^{(t)}+iY_\nu^{(t)}$, where 
 $X_\nu^{(t)}$ and $Y_\nu^{(t)}$ are Hermitian. We then have 
\be
D(A^{(t)}_\nu)^\dagger & = & (\gamma_\mu(\partial_\nu+i(X^{(t),a}_\nu+iY^{(t),a}_\nu)\lambda^a+\mu\delta_{\nu,0}))^\dagger \nn\\
&=& -\gamma_\mu(\partial_\nu+i(X^{(t),a}_\nu-iY^{(t),a}_\nu)\lambda^a-\mu\delta_{\nu,0}), \nn \\
\ee 
where the $\lambda^a$ are the color generators of SU(3). To make the full Dirac operator anti-Hermitian would require 
that the $Y_\nu^{(t),a}$ nulify the chemical potential and this is not possible since the unit matrix in color is not a generator of SU(3).
A total anti-Hermitization  of the Dirac operator by complixification of the gauge fields is therefore not possible \footnote{Note that with 
gauge group U(3) the Dirac operator at non-zero $\mu$ can be absorbed into a global shift of the gauge fields in the complex plane.  
Complex Langevin simulations for QCD with gauge group U(3) have been reported to be successful \cite{Karsch:1987ry}.}. 
The new constraints on the spectrum found above, however, only require that the small Dirac eigenvalues accumulate at the origin in the chiral limit. 
For this a total anti-Hermitization of the Dirac operator is not required.

The fact that the quark mass hits the spectrum of the Dirac operator at $\mu=m_\pi/2$ for real gauge fields is not accidental:
This follows  directly from the spontaneous breakdown of chiral symmetry and the fact that the complex conjugate of the 
fermion determinant equals the determinant with the opposite sign of the chemical potential \cite{Misha,TV,OSV-phaseSpec}.
This standard argument does not exclude the possibility that the spectral density of the Dirac operator
with complexified gauge fields remains inside the quark mass even if $\mu>m_\pi/2$:
In the case of complex Langevin the complex conjugate of the fermion determinant is
\be
&&\det(\gamma_\mu(\partial_\nu+i(X^{(t),a}_\nu+iY^{(t),a}_\nu)\lambda^a+\mu\delta_{\nu,0})+m)^*\nn\\
&=&\det(\gamma_\mu(\partial_\nu+i(X^{(t),a}_\nu-iY^{(t),a}_\nu)\lambda^a-\mu\delta_{\nu,0})+m).\nn 
\ee
For non-zero values of $Y^{(t)}_\nu$ (i.e.~when the gauge field is complex) this is not equal to the 
determinant at $-\mu$. Hence spontaneous breaking of chiral symmetry 
does not automatically imply that the quark mass hits the support of the Dirac spectrum with complexified 
gauge fields for $\mu=m_\pi/2$. 
The necessary behaviour of the small Dirac eigenvalues in complex Langevin simulations sketched in the lower 
panels of Figure \ref{fig:fixed-m} and \ref{fig:fixed-mu} and exemplified in the appendix, is therefore not excluded 
by the standard argument.

\begin{center}
\begin{figure}[t!]
\includegraphics[width=8cm,angle=0]{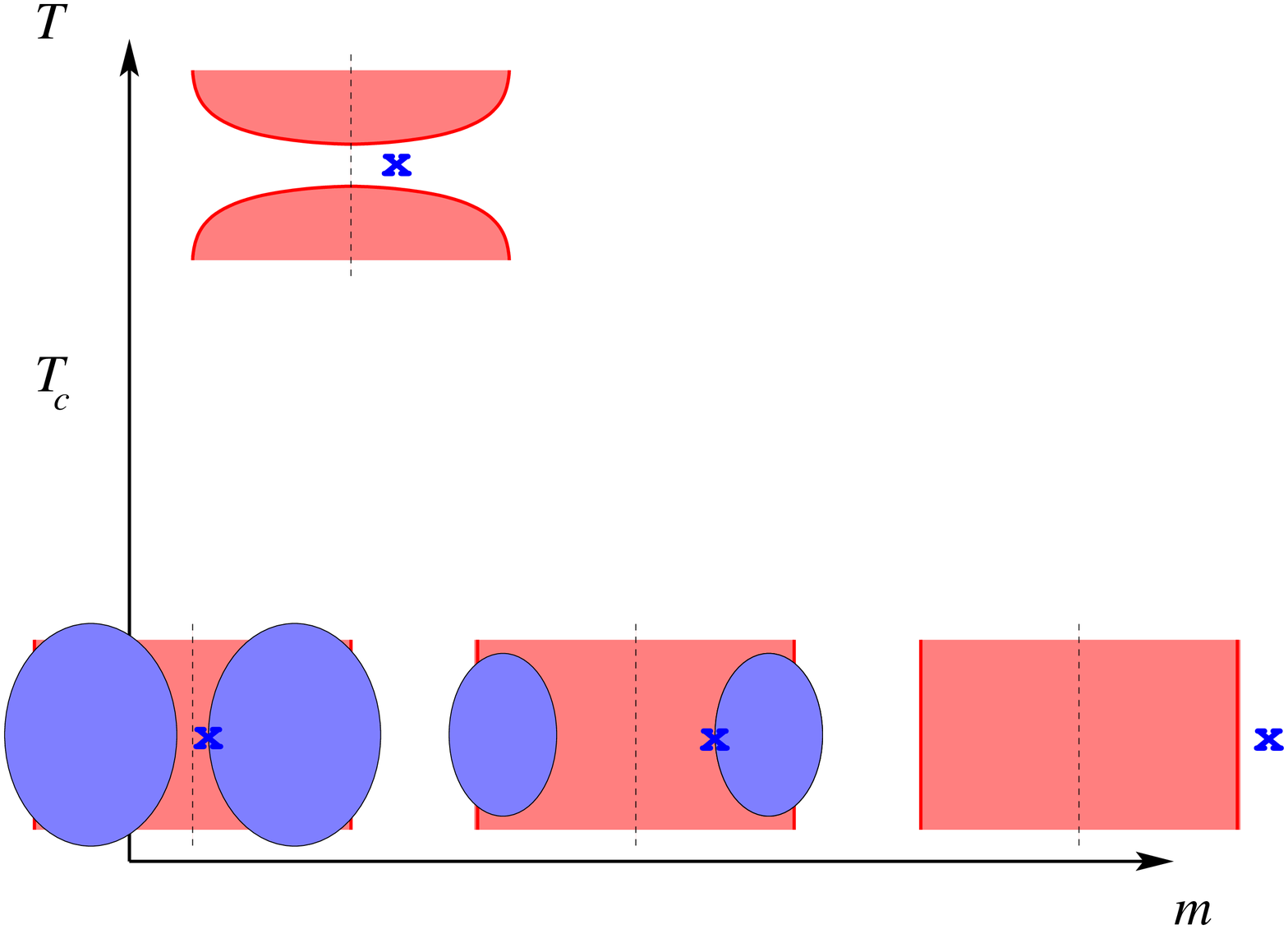}
\vspace{10mm}
\vfill
\includegraphics[width=8cm,angle=0]{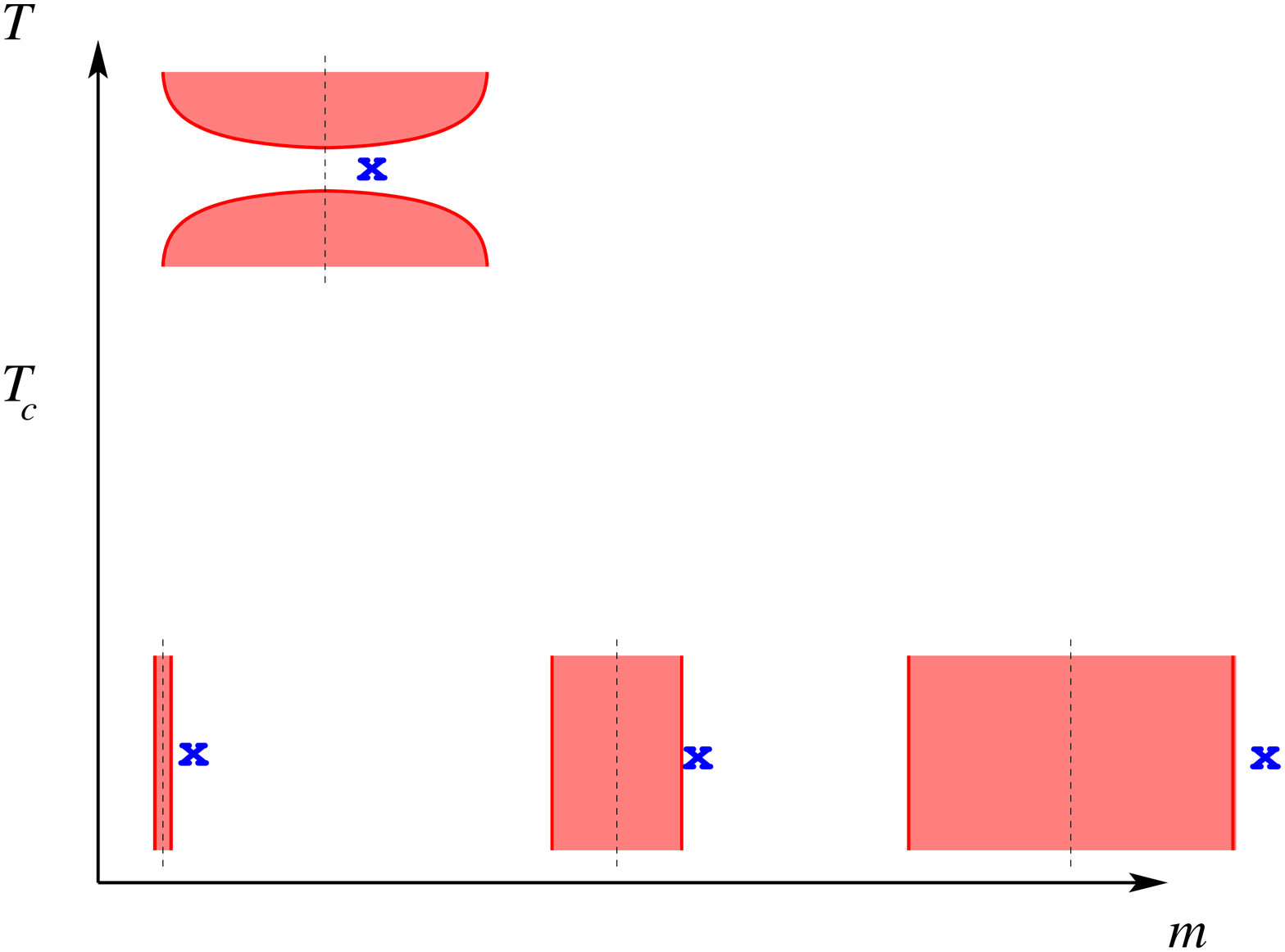}
\caption{\label{fig:fixed-mu} The schematic behaviour of the low lying Dirac spectrum in the $(m,T)$-plane at fixed non-zero $\mu$ (where $|\mu|$ is less than a third of the nucleon mass). {\bf Top:} with real gauge fields, {\bf Below:} with complexified fields from complex Langevin. The eigenvalue density is real and positive within the red region and complex within the blue. For real gauge fields the quark mass (marked by the blue X) 
hits the eigenvalue density when $m_\pi=2\mu$. In complex Langevin simulations the eigenvalues must be inside the quark mass.}\vspace{5mm}
\end{figure}
\end{center}

\subsection{The cut of the log and a practical suggestion for simulations}

As demonstrated in \cite{MS} the presence of the log of the fermion determinant in the action may 
lead to inconsistencies if the determinant frequently circles the origin during 
the Langevin flow, see also \cite{Greensite}. These inconsistencies arise because the complex Langevin drift 
includes the derivative of the non-holomorfic logarithm in the complex plane. This becomes 
particularly relevant when the argument of the log passes the cut of the log.

If we express the log of the fermion determinant through the eigenvalues of the Dirac operator, 
\be
\log\det(D_\mu(A_\nu^{(t)})+m)=\sum_k\log (z_k^{(t)}+m),
\ee
the potential problems with the log become relevant when the quark mass is inside 
the support of the eigenvalues, such that $z_k^{(t)}+m$ or $-z_k^{(t)}+m$ can circle the origin of the 
complex plane (recall that for every non-zero $z_k^{(t)}$ there is an eigenvalue $-z_k^{(t)}$). 
Now, as we have seen above, in order that complex Langevin correctly describes the spontaneous 
breaking of chiral symmetry, at low temperature and $\mu$ up to a third of the nucleon mass, the small
eigenvalues must end up inside the quark mass. If the eigenvalues indeed do so the problems 
with the log will be minimal and complex Langevin can be a solution to the QCD sign problem.

If the complex Langevin eigenvalues end up inside the quark mass there still remains, a 
problem of 'thermalization': 
If at Langevin time, $t=1$, the trajectory starts from a typical (i.e.~a quenched, a $\mu=0$, 
or a phase quenched) real gauge field configuration the quark mass will initially be inside the support 
of the Dirac eigenvalues if $\mu>m_\pi/2$. Therefore with Langevin time the eigenvalues must
must flow inside the quark mass. However, in order to safely ignore the issues with the log, the 
Langevin flow of the Dirac eigenvalues must be such that the eigenvalues do not circle the quark mass. 
Even if the Langevin drift may not wish to drive the eigenvalues around the quark mass  
\footnote{Trajectories where the imaginary part of the action remain constant and which end up at 
fixed points are known as Lefschetz thimbles \cite{Lef}. In \cite{AartsCL-vs-L,ABSS} it was demonstrated that  
in some specific examples, the regions of the complex plane sampled by 
complex Langevin are in the neighbourhood of the Lefschets thimble. If this is true in full QCD  
then the determinant does not frequently circle the origin.}
it will take a very careful update procedure (a small time step) to avoid that this happens 
due to the Langevin noise: If the quark mass initially is inside the support of the Dirac eigenvalues 
then the typical distance from the quark mass to the nearest eigenvalue is of order the inverse 
volume \cite{AOSV} and a numerical fluctuation can cause this eigenvalue to move to the other side of the quark mass.  
In practice, it is also exceedingly demanding to invert the Dirac operator numerically when the quark
mass is inside the cloud of eigenvalues. 

In order to overcome this we here {\sl suggest initially to let the quark mass 
depend on the Langevin time}. If the Langevin trajectory starts from a typical real gauge field configuration 
and we pick a quark mass such that it is close to but outside the eigenvalues 
of the Dirac operator on the original configuration, then we can run the Langevin flow ignoring the 
issues with the cut of the log. If at a suitable Langevin time $t_1$ the small eigenvalues of the Dirac operator evaluated 
on the latest complex valued configuration is in a narrower band than at time $t=1$ we then decrease the 
quark mass towards the desired value but no further than it is still outside the band of Dirac eigenvalue at $t_1$.
 Continuing in this manner one may potentially take the quark mass all the way to 
the desired value without ever having the quark mass inside the support of the eigenvalues. Once the 
desired value of the quark mass is reached, all prior configurations are discarded in the Langevin measurement. 
Alternatively one may work at a fixed quark mass and increase the chemical potential  with Langevin time until the 
desired value is reached \footnote{I thank Prof.~F.~Karsch for this suggestion given during a seminar at University of Bielefeld 
December 2013.}.

\section{Conclusions}

We have considered, for the first time, the general properties of the Dirac eigenvalues on the complexified 
gauge field configurations generated by complex Langevin.  By the very definition of 
the expectation value  in complex Langevin the density of Dirac eigenvalues on the 
Langevin trajectory of complexified gauge fields will be real and positive. 
Therefore, using complex Langevin simulations it is not possible to realize the complex and 
oscillating eigenvalues density which is essential for the spontaneous breaking of chiral 
symmetry with real gauge fields. 
The eigenvalues on the complexified gauge fields must therefore
distribute themselves in a manner which is drastically different from the distribution obtained with 
real gauge fields: The small Dirac eigenvalues, when evaluated on the 
Langevin trajectory, must form a band inside the quark mass. In particular, in the chiral limit 
the small eigenvalues  must accumulate at the origin. 
  
A non-positive Dirac eigenvalue density where oscillations with a period of order the inverse volume and 
an exponentially large amplitude is a generic feature of theories with a fermionic sign problem. It is
essential in the microscopic domain of QCD \cite{OSV1,OSV2,IS}, for Cooper paring at large chemical 
potential \cite{KanazawaWettig}, in one dimensional QED at any chemical potential \cite{RV,AS}, in
two color QCD with unmatched quark masses \cite{AKPW}, as 
well as for odd flavored QCD at zero chemical potential and a negative quark mass \cite{VerbWettig}.  
All cases are characterized by a fermonic sign problem and the results obtained here also applies in 
these cases: If complex Langevin is applied the resulting Dirac spectrum must be drastically different 
from the one obtained with real fields.

Finally, we have proposed to work with an initially Langevin time dependent quark mass. It would 
be most interesting to implement this in complex Langevin simulations of full QCD. A first test 
in the context of chiral random matrix theory \cite{O,canonical} is presented in \cite{MS2}.

\vspace{5mm}

\noindent
{\bf Acknowledgments:}  Jac Verbaarschot, Poul Henrik Damgaard, Mario Kieburg, Gernot Akemann, Philippe de Forcrand, Frithjof Karsch, 
Gert Aarts, Jeff Greensite, Francesco Di Renzo, Erhard Seiler, Christian Schmidt, Denes Sexty, Nucu Stamatescu, Luigi Scorzato, Joyce C.~Myers 
and Anders Mollgaard as well as participants of Sign 2014, XQCD 2014 and 'Conceptual advances in lattice 
gauge theory (LTG14 at CERN)' are thanked for discussions. The work of KS was supported by the {\sl Sapere Aude program} of The Danish 
Council for Independent Research. 


\renewcommand{\thesection}{\Alph{section}}
\setcounter{section}{0}

\section{Three examples}
\label{app:3ex}

We here give three examples of positive spectral densities and discuss the resulting chiral condensates.
\vspace{2mm}

{\bf 1) a uniform strip of width $2\mu^2F_\pi^2/\Sigma$} 
\vspace{2mm}

The first example is the positive eigenvalue density obtained in phase quenched QCD. At mean field level in 
chiral perturbation theory \cite{TV,SV-fact} it forms a strip around the imaginary axis of width 
$2\mu^2F_\pi^2/\Sigma$. Within this strip the density is constant  
\be
\rho(z,m;\mu) = \frac{\Sigma^2}{4\pi\mu^2F_\pi^2}\;\theta(2\mu^2F_\pi^2/\Sigma-|x|),
\ee
where $z=x+iy$. Note that the eigenvalue density is independent of the quark mass and that the normalization 
of the density requires that the density itself diverges as $\mu\to0$. 

The resulting chiral condensate, 
\be
\Sigma(m;\mu)& =& \frac{\Sigma^2}{2\mu^2F_\pi^2} m \; \theta(2\mu^2F_\pi^2/\Sigma-|m|)\\ 
&&\hspace{-3mm}+\; \Sigma \; {\rm sign}(m)\; \theta(-2\mu^2F_\pi^2/\Sigma+|m|), \nn
\ee
has no discontinuity as $m$ passes through zero. Moreover, the kink at $m=2\mu^2F_\pi^2/\Sigma$ 
is the signal of the second order phase transition at $\mu=m_\pi/2$ as follows using the 
GOR relation $m_\pi^2F_\pi^2=2\Sigma m$.
\vspace{2mm}

{\bf 2) a uniform strip of width $|m|/2$} 
\vspace{2mm}

In full QCD with real gauge fields the eigenvalue density is complex and depends heavily on 
the quark mass $m$. The Dirac eigenvalue density in complex Langevin simulations of full QCD is real 
and positive definite by construction and the support must be inside the quark mass for low temperature. To exemplify 
this we consider here an example of a positive eigenvalue density which is in a strip of width $|m|/2$ 
\be
\rho(z,m;\mu) = \frac{\Sigma}{\pi|m|}\theta(|m|/2-|x|),
\ee
where $x={\rm Re}[z]$. 
In order to make the example more realistic let us assume that this is true for $|m|<4\mu^2F_\pi^2/\Sigma$
while for larger values of $|m|$ the density is given as in the first example above. 

Note that this eigenvalue density diverges as $m\to0$ in order to keep the normalization fixed (it becomes a 
one-dimensional density on the imaginery axis for $m=0$). This leads to a discontinuous chiral condensate, 
\be
\Sigma(m) & = & \Sigma\;{\rm sign}(m) ,
\ee
signalling the spontaneous breakdown of 
chiral symmetry in the chiral limit.

The diverging density at the origin in the chiral limit exemplifies the behaviour which the eigenvalue density must have in 
complex Langevin simulations in order to support the spontaneous breaking of chiral symmetry. 
\vspace{2mm}

{\bf 3) scaling relation} 
\vspace{2mm}

It is tempting to come up with examples where the eigenvalue density for a real number $k$ satisfies the scaling relation
\be
\rho(kz,km;\mu) = \frac{1}{|k|} \rho(z,m;\mu).
\label{scaling}
\ee
For example
\be
\rho(z,m;\mu) = \frac{\Sigma}{2|y| + |m|}\frac{1}{\pi}\; \theta(|y| + |m/2| - |x|),
\ee
where $z=x+iy$ so that the support has the form of an hourglass. The minimal width $|m|/2$ is obtained at $y=0$ and from there the width 
grows strictly linearly with $|y|$. 

With the scaling relation (\ref{scaling}) we automatically get that 
\be
\Sigma(km) & = & \int dxdy \frac{1}{x+iy+km} \rho(z,km;\mu) \\
                   & = & k^2 \int dx'dy' \frac{1}{kx'+iky'+km} \rho(kz',km;\mu) \nn \\
                   & = & \frac{k^2}{k|k|} \int dx'dy' \frac{1}{x'+iy'+m} \rho(z',m;\mu) \nn \\
                   & = & {\rm sign}(k) \Sigma(m),\nn
\ee
where $z=x+iy$ and $z=kz'$. As desired the condensate is $m$ independent except that it changes sign at $m=0$. There is a 
catch, however, since the scaling relation (\ref{scaling}) is inconsistent with the fact that the normalization of the eigenvalue density, 
\be
\int d^2z \ \rho(z,m;\mu),
\ee
must be independent of $m$. For this to hold true the value of $k$ in (\ref{scaling}) must be 1 and the scaling relation is therefore trivial.




\begin{thebibliography}{99}



\bibitem{Parisi} 
  G.~Parisi,
  Phys.\ Lett.\ B {\bf 131}, 393 (1983).

\bibitem{Klauder} 
  J.~R.~Klauder,
  J.\ Phys.\ A {\bf 16}, L317 (1983).

\bibitem{AartsBose} 
  G.~Aarts,
  Phys.\ Rev.\ Lett.\  {\bf 102}, 131601 (2009)
  [arXiv:0810.2089 [hep-lat]].

\bibitem{AS} 
  G.~Aarts and K.~Splittorff,
  JHEP {\bf 1008}, 017 (2010)
  [arXiv:1006.0332 [hep-lat]].

  \bibitem{AFP} 
  J.~Ambjorn, M.~Flensburg and C.~Peterson,
  Nucl.\ Phys.\ B {\bf 275}, 375 (1986).
  
  \bibitem{AartsXY} 
  G.~Aarts and F.~A.~James,
  JHEP {\bf 1008}, 020 (2010)
  [arXiv:1005.3468 [hep-lat]].
  
  \bibitem{MS} 
  A.~Mollgaard and K.~Splittorff,
  Phys.\ Rev.\ D {\bf 88}, 116007 (2013)
  [arXiv:1309.4335 [hep-lat]].


\bibitem{AartsRev} 
  G.~Aarts,
  PoS LATTICE {\bf 2012}, 017 (2012)
  [arXiv:1302.3028 [hep-lat]].


\bibitem{PhilippeRev} 
  P.~de Forcrand,
  PoS LAT {\bf 2009}, 010 (2009)
  [arXiv:1005.0539 [hep-laxt]].

\bibitem{GaugeCooling} 
  E.~Seiler, D.~Sexty and I.~O.~Stamatescu,
  Phys.\ Lett.\ B {\bf 723}, 213 (2013)
  [arXiv:1211.3709 [hep-lat]].

\bibitem{Sexty:2013ica} 
  D.~Sexty,
  Phys.\ Lett.\ B {\bf 729}, 108 (2014)
  [arXiv:1307.7748 [hep-lat]].

\bibitem{all-orders} 
  G.~Aarts, E.~Seiler, D.~Sexty and I.~O.~Stamatescu,
  arXiv:1408.3770 [hep-lat].

\bibitem{SVphase} 
  K.~Splittorff and J.~J.~M.~Verbaarschot,
  Phys.\ Rev.\ Lett.\  {\bf 98}, 031601 (2007)
  [hep-lat/0609076];
  Phys.\ Rev.\ D {\bf 75}, 116003 (2007)
  [hep-lat/0702011 [HEP-LAT]].

\bibitem{SplitRev} 
  K.~Splittorff,
  PoS LAT {\bf 2006}, 023 (2006)
  [hep-lat/0610072].

\bibitem{TV} 
  D.~Toublan and J.~J.~M.~Verbaarschot,
  Int.\ J.\ Mod.\ Phys.\ B {\bf 15}, 1404 (2001)
  [hep-th/0001110].

\bibitem{AOSV} 
  G.~Akemann, J.~C.~Osborn, K.~Splittorff and J.~J.~M.~Verbaarschot,
  Nucl.\ Phys.\ B {\bf 712}, 287 (2005)
  [hep-th/0411030].

\bibitem{SS} 
  D.~T.~Son and M.~A.~Stephanov,
  Phys.\ Rev.\ Lett.\  {\bf 86}, 592 (2001)
  [hep-ph/0005225].
  
\bibitem{SSS} 
  K.~Splittorff, D.~T.~Son and M.~A.~Stephanov,
  Phys.\ Rev.\ D {\bf 64}, 016003 (2001)
  [hep-ph/0012274].

\bibitem{OSV-phaseSpec} 
  J.~C.~Osborn, K.~Splittorff and J.~J.~M.~Verbaarschot,
  Phys.\ Rev.\ D {\bf 78}, 105006 (2008)
  [arXiv:0807.4584 [hep-lat]].

\bibitem{OSV1} 
  J.~C.~Osborn, K.~Splittorff and J.~J.~M.~Verbaarschot,
  Phys.\ Rev.\ Lett.\  {\bf 94}, 202001 (2005)
  [hep-th/0501210].

\bibitem{OSV2} 
  J.~C.~Osborn, K.~Splittorff and J.~J.~M.~Verbaarschot,
  Phys.\ Rev.\ D {\bf 78}, 065029 (2008)
  [arXiv:0805.1303 [hep-th]].

\bibitem{BC} 
  T.~Banks and A.~Casher,
  Nucl.\ Phys.\ B {\bf 169}, 103 (1980).

\bibitem{LS} 
  H.~Leutwyler and A.~V.~Smilga,
  Phys.\ Rev.\ D {\bf 46}, 5607 (1992).

\bibitem{OSV-XQCD} 
  J.~C.~Osborn, K.~Splittorff and J.~J.~M.~Verbaarschot,
  hep-lat/0510118.

\bibitem{O} 
  J.~C.~Osborn,
  Phys.\ Rev.\ Lett.\  {\bf 93}, 222001 (2004)
  [hep-th/0403131].

\bibitem{Misha} 
  M.~A.~Stephanov,
  Phys.\ Rev.\ Lett.\  {\bf 76}, 4472 (1996)
  [hep-lat/9604003].

 \bibitem{Greensite} 
  J.~Greensite,
  arXiv:1406.4558 [hep-lat].

 \bibitem{IS} 
  J.~R.~Ipsen and K.~Splittorff,
  Phys.\ Rev.\ D {\bf 86}, 014508 (2012)
  [arXiv:1205.3093 [hep-lat]].
  
   \bibitem{KanazawaWettig} 
  T.~Kanazawa and T.~Wettig,
  arXiv:1406.6131 [hep-ph].

\bibitem{RV} 
  L.~Ravagli and J.~J.~M.~Verbaarschot,
  Phys.\ Rev.\ D {\bf 76}, 054506 (2007)
  [arXiv:0704.1111 [hep-th]].

\bibitem{AKPW} 
  G.~Akemann, T.~Kanazawa, M.~J.~Phillips and T.~Wettig,
  JHEP {\bf 1103}, 066 (2011)
  [arXiv:1012.4461 [hep-lat]].
 

  \bibitem{VerbWettig} 
  J.~J.~M.~Verbaarschot and T.~Wettig,
  arXiv:1407.8393 [hep-th].

  \bibitem{canonical} 
  J.~Bloch, F.~Bruckmann, M.~Kieburg, K.~Splittorff and J.~J.~M.~Verbaarschot,
  Phys.\ Rev.\ D {\bf 87}, no. 3, 034510 (2013)
  [arXiv:1211.3990 [hep-lat]].

   \bibitem{MS2}
A. ~Mollgaard and K.~Splittorff, {\sl in preparation}. 
 
  \bibitem{SV-fact} 
  K.~Splittorff and J.~J.~M.~Verbaarschot,
  Nucl.\ Phys.\ B {\bf 683}, 467 (2004)
  [hep-th/0310271].

  
  \bibitem{Karsch:1987ry} 
  F.~Karsch,
  Nucl.\ Phys.\ A {\bf 461}, 305C (1987).
  
    
  \bibitem{KSV} 
  M.~Kieburg, K.~Splittorff and J.~J.~M.~Verbaarschot,
  Phys.\ Rev.\ D {\bf 85}, 094011 (2012)
  [arXiv:1202.0620 [hep-lat]].
   
   \bibitem{Lef} 
  M.~Cristoforetti {\it et al.}  [AuroraScience Collaboration],
  Phys.\ Rev.\ D {\bf 86}, 074506 (2012)
  [arXiv:1205.3996 [hep-lat]].
  
  \bibitem{AartsCL-vs-L} 
  G.~Aarts,
  Phys.\ Rev.\ D {\bf 88}, no. 9, 094501 (2013)
  [arXiv:1308.4811 [hep-lat]].
  
  \bibitem{ABSS} 
  G.~Aarts, L.~Bongiovanni, E.~Seiler and D.~Sexty,
  arXiv:1407.2090 [hep-lat].
    
   
     
  
\end{thebibliography}
\end{document}